%% file: ms.tex
\documentclass[12pt,preprint]{aastex}

\begin{document}

\title{The spatial distribution of stars in open clusters}

\author{N\'estor S\'anchez\altaffilmark{1} and
        Emilio J. Alfaro\altaffilmark{1}}

\altaffiltext{1}{Instituto de Astrof\'{\i}sica de Andaluc\'{\i}a,
                 CSIC, Apdo. 3004, E-18080, Granada, Spain.}

\slugcomment{The Astrophysical Journal: accepted}

\begin{abstract}
The analysis of the distribution of stars in open clusters may
yield important information on the star formation process and
early dynamical evolution of stellar clusters. Here we address
this issue by systematically characterizing the internal spatial
structure of 16 open clusters in the Milky Way spanning a wide 
range of ages. Cluster stars have been selected from a membership
probability analysis based on a non-parametric method that uses both
positions and proper motions and does not make any a priori assumption
on the underlying distributions. The internal structure is then
characterized by means of the minimum spanning tree method ($Q$
parameter), King profile fitting, and the correlation dimension
($D_c$) for those clusters with fractal patterns.
On average, clusters with fractal-like structure are younger than
those exhibiting radial star density profiles and an apparent
trend between $Q$ and age is observed in agreement with previous
ideas about the dynamical evolution of the internal spatial
structure of stellar clusters. However, some new results are
obtained from a more detailed analysis: (a) a clear correlation
between $Q$ and the concentration parameter of the King model for
those cluster with radial density profiles, (b) the presence of
spatial substructure in clusters as old as $\sim 100$ Myr, and
(c) a significant correlation between fractal dimension and age
for those clusters with internal substructure. Moreover, the
lowest fractal dimensions seem to be considerably smaller than
the average value measured in galactic molecular cloud complexes.
\end{abstract}

\keywords{ISM: structure ---
          methods: statistical ---
          open clusters and associations: general ---
          stars: formation}


\section{Introduction}

It is known that most stars are born within giant molecular clouds 
forming clusters \citep{Lad03}. Numerical simulations demonstrate
that star formation occurs mainly along the patterns defined by the
densest regions of the molecular clouds \citep{Bon03}. Thus, the
hierarchical structure observed in some open clusters \citep[see,
for example,][]{Lar95} is presumably a consequence of its formation
in a medium with an underlying fractal structure. This fractality
is considered to be a clear signature of its own turbulent nature
\citep{Elm04}. Otherwise, open clusters having central star
concentrations with radial star density profiles likely reflect
the dominant role of gravity, either on the primordial gas
structure or as a result of a rapid evolution from a more
structured state \citep{Lad03}.

It is therefore important to study the distribution of stars
because it may yield some information on the formation process
and early evolution of open clusters. It is necessary, however,
that this kind of analysis is done by measuring the cluster
structure in an objective, quantitative, as well as systematic
way. The application of the two-point correlation function by
\citet{Lar95} to young stars in Taurus suggested that the
distribution of stars on spatial scales larger than the binary
regime exhibits a fractal pattern with a projected dimension of
$D \sim 1.4$. This value is very similar to the average value
$D \sim 1.5 \pm 0.2$ found by \citet{Sim97} for the Taurus,
Ophiuchus, and Orion trapezium regions. More recent works find
significantly smaller values for stars in Taurus,
such as $D = 1.02 \pm 0.04$ \citep{Har02} and $D =
1.049 \pm 0.007$ \citep{Kra08}. The difference could be at
least partly due to differences in the completeness of the
sample \citep{San08}. \citet{Nak98} studied the clustering
of stars in the Orion, Ophiuchus, Chamaeleon, Vela, and
Lupus regions obtaining significant variations from region
to region in the range $1.2 < D < 1.9$. The interpretation 
of these results requires some caution because it has been
shown that a good power-law fit to the two-point correlation 
function does not necessarily mean that the stellar
distribution is fractal \citep{Bat98}.

\citet{Car04} developed a different method
to quantify the structure of star clusters.
Their method is based on the construction of
the minimum spanning tree (MST) of the cluster and it has the
important advantage of being able to distinguish between
centrally concentrated and fractal-like structures. They
concluded that the fractal dimension of the three-dimensional
distribution of stars in Chamaeleon and IC~2391 is
$D_f = 2.25 \pm 0.25$, whereas in Taurus $D_f = 1.9 \pm 0.2$
\citep[see also][]{Sch06}. These dimensions seem too small when
compared with the average value of $D_f \simeq 2.6-2.7$ suggested
for the structure of the interstellar medium \citep{San05,San07b},
although higher dimension values have been reported using this
technique (MST) on clusters in other regions such as Serpens
and Ophiuchus \citep{Sch08}. The rapid early evolution of star
clusters may complicate the picture, because the parameters
characterizing the cluster structure must only be taken as
instantaneous values which might change
significantly in a few Myr
\citep{Bas08}. \citet{Sch06} applied the MST method to both
observed and simulated clusters to argue that star clusters
preferentially form with a clustered, fractal-like structure
and gradually evolve to a more centrally concentrated state
\citep[see also][]{Sch08}. In any case, some kind of
relationship between the {\it initial} structure of the
clusters and the properties of the turbulent medium where
they were born is expected \citep{Bal07}. \citet{Sch08}
find certain evidence that regions with relatively high
Mach numbers form clusters more hierarchically structured,
i.e. with relatively small fractal dimensions. They estimated
a Mach number of $M \simeq 5.8$ in the Ophiuchus region where
the cluster L1688 is found, for which they reported a structure
parameter compatible with $D_f \sim 2.5$. These results would
be in agreement with simulations of turbulent fragmentation
in molecular clouds \citep{Bal06}, but it has to be pointed out
that other studies do not find such a correlation \citep{Sch06}
and others directly contradict it \citep{Eno07}. \citet{Fed07}
used numerical simulations of supersonic turbulence to show
that, for the same Mach number ($M \simeq 5.5$), the fractal
dimension of the medium can be very different ranging from
$D_f \simeq 2.3$ to $D_f \simeq 2.6$ depending on whether
turbulence is driven by the usually adopted solenoidal
forcing or by compressive forcing, respectively.

In this work, we consider this subject by systematically
analyzing the distribution of stars in a sample of open
clusters spanning a representative range of age and distance
values with kinematical data available in the literature.
The clusters are visible at optical
wavelengths possibly indicating that even the youngest
ones have dispersed most of the gas and dust from which
they were born. Obviously, these objects may present
significant contamination by field stars projected
along the line of sigh. The MST technique tends to
lose information on the degree of fractality as the number
of contaminating field stars increases \citep{Bas09}.
Moreover, and very important, the combination of data
coming from different sources with different membership
selection criteria might introduce undesired scatter
as well as some bias in the final results.
To overcome these problems, we decided to calculate the
memberships by applying the same general, non-parametric
method to all the clusters. In order to achieve
a representative work sample (Section~\ref{membresias}),
we first collect in Section~\ref{sample} as much data as
possible on positions and proper motions of stars in open
cluster regions. Using these data, we apply in Section~\ref{salson}
the non-parametric method to assign cluster memberships.
A comparison between these memberships and those
obtained from the classical parametric method is
done in Section~\ref{comparacion}.
The distribution of the stars is then quantified in
Section~\ref{resultados} by means of the MST technique,
King profile fittings, and the correlation dimension if
the distribution is fractal. The dependence of the cluster
structure on its age is discussed in Section~\ref{correlaciones}.
Finally, the main results are summarized in Section~\ref{conclusiones}.


\section{Star cluster membership}
\label{membresias}

\subsection{The sample of clusters}
\label{sample}

We first used VizieR\footnote{http://vizier.u-strasbg.fr}
\citep{Och00} to search for catalogs of open clusters containing
both positions and proper motions available in machine-readable
format. We required the data to be available for all the stars
in the field and not only for the probable members according
to each author's criteria. Then we checked the catalogs and
rejected those that could generate some sort of bias. For
example, catalogs containing data only for a specified region
of the cluster or for a limited sample of stars were ruled out.
In the end, we have a total of 16 open clusters which are listed
in Table~\ref{cumulos}. This
table also gives the logarithm of the cluster age in Myr
($\log (T)$) and the distance in pc ($D$), taken from the
Webda database\footnote{http://www.univie.ac.at/webda}, as
well as the number of stars having positions and proper
motions in the original catalog ($N_d$), the number of stars
selected as cluster members in Section~\ref{salson} ($N_s$),
and the values calculated in Section~\ref{resultados} for the
structure parameter ($Q$) and the core ($R_c$) and tidal ($R_t$)
radii in pc. The last column in Table~\ref{cumulos} lists the
references from which the data used in this work were taken.
We have to mention that clusters in this sample have been
observed at optical wavelengths. They have little or no
primordial interstellar gas in them and therefore they
may be in a supervirialized state \citep{Goo04}, mainly
the youngest ones.

\subsection{Non-parametric method}
\label{salson}

An initial step in any study on open clusters is the reliable
identification of probable members. This is a complex 
problem that deserves to be addressed more comprehensively.
Several different methods for estimating membership
probabilities may be used depending on whether one is dealing
with positions, proper motions, radial velocities, multiband
photometry, or a combination of them. However, it is commonly
accepted that membership probabilities obtained from kinematical
variables are more reliable than those derived from other
kind of physical variables. When working with proper motion
data, the most often used method is the algorithm designed by
\citet{San71} based on the former model proposed by \citet{Vas58}
for the proper motion distribution in the cluster vicinity.
The method assumes that the two populations (cluster
members and field stars) are distributed according to normal
bivariate functions and then the observed distribution is
a weighted mixture of these two underlying
distributions. It can be proven that the
classification and estimation problem derived from this model
has mathematical solutions. Some problems may arise when 
applying the method of \citet{San71} if the two parent 
populations are very far from the mathematical
functions on which the model is based \citep{Pla01}.
In order to prevent this and other potential problems, \citet{Cab90}
developed a more general, non-parametric method which makes no
{\it a priori} assumptions about the cluster and field star 
distributions. Besides the proper motions, the method uses
the spatial distribution of stars as a complementary
and necessary source of information.
Generally speaking, the method iteratively estimates the
probability density function using Kernel functions with smooth
parameters such that the likelihood is maximum \citep[see
details in][]{Cab90}. The only astronomical hypotheses
remaining are that there are two populations (cluster
and field) and that cluster members are more densely
distributed than field stars (both in proper motions
and in positions). An important distinction between the
classical Sanders method and this one is that here the
classification of the stars can be done
according to three different probabilities: the probability
derived from the position space $P(x,y)$, the probability
derived from the proper motion space $P(\mu_x,\mu_y)$, and
the joint probability $P(x,y,\mu_x,\mu_y)$.

We applied this method to our sample of open cluster
(Section~\ref{sample}) in a systematic and self-consistent
way. We used exactly the same algorithm and the same
selection criteria for cluster members:
$P(x,y,\mu_x,\mu_y) \ge 0.5$ and $P(\mu_x,\mu_y) \ge 0.5$.
This choice puts more weight on the kinematical variables than
on the positional variables. If the algorithm did not find
any cluster member (this happened in 5 of the 16 cases)
then the joint probability criterion was changed to
$P(x,y,\mu_x,\mu_y) \ge 0.4$ but no additional condition
was needed to achieve convergence.
It is worth noting that, given the iterative nature of
this method, the final membership probabilities are in
principle dependent upon the decision rule chosen. We
have performed some tests by varying the selection
criteria around the above values and, although there
were changes in the membership assignments, the results
and trends obtained on the spatial structure of the
clusters (next sections) remained practically unaltered.

\begin{figure}[th]
\epsscale{.9}
\plotone{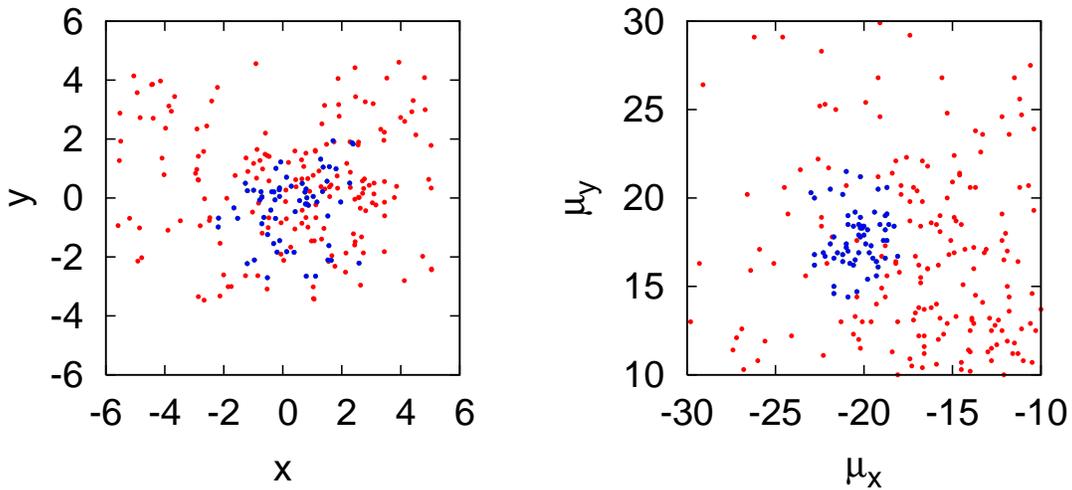}
\caption{Positions ({\it left}) in pc relative to the center
and proper motions ({\it right}) in mas/yr for the stars in
the region of the open cluster IC~2391. Red circles indicate
field stars and blue circles cluster members according to
the method applied in this work.}
\label{ic2391}
\end{figure}

\begin{figure}[th]
\epsscale{.9}
\plotone{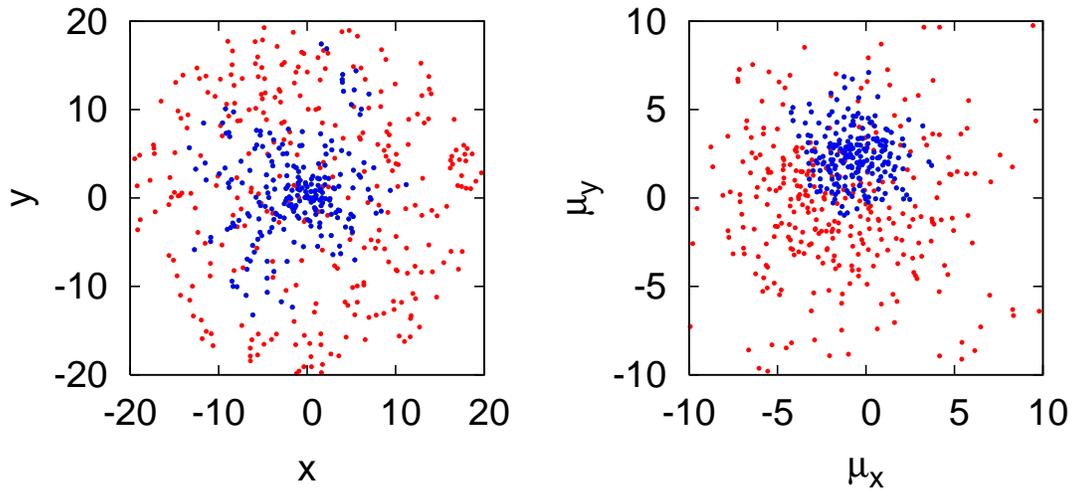}
\caption{Same as Fig.~\ref{ic2391}, but for the open
cluster NGC~2194.}
\label{ngc2194}
\end{figure}

\begin{figure}[th]
\epsscale{.9}
\plotone{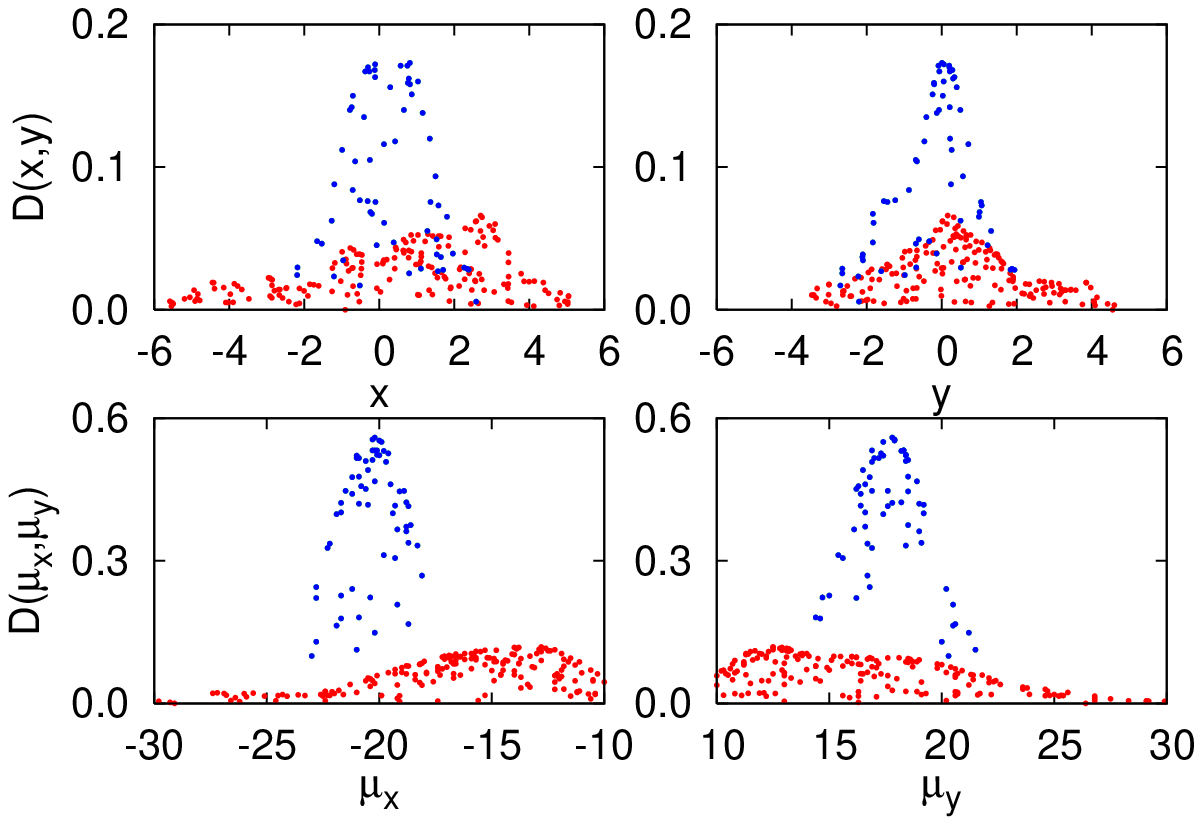}
\caption{Probability density funcions for the stars in
the region of the open cluster IC~2391. The two upper panels
show the projections in $x$ and $y$ of the probability
densities in the position space. The two lower panels
are the projections in $\mu_x$ and $\mu_y$  of the
probability densities in the proper motion space. Red
circles refer to field stars and the blue ones to cluster
members.}
\label{pdfic2391}
\end{figure}

\begin{figure}[th]
\epsscale{.9}
\plotone{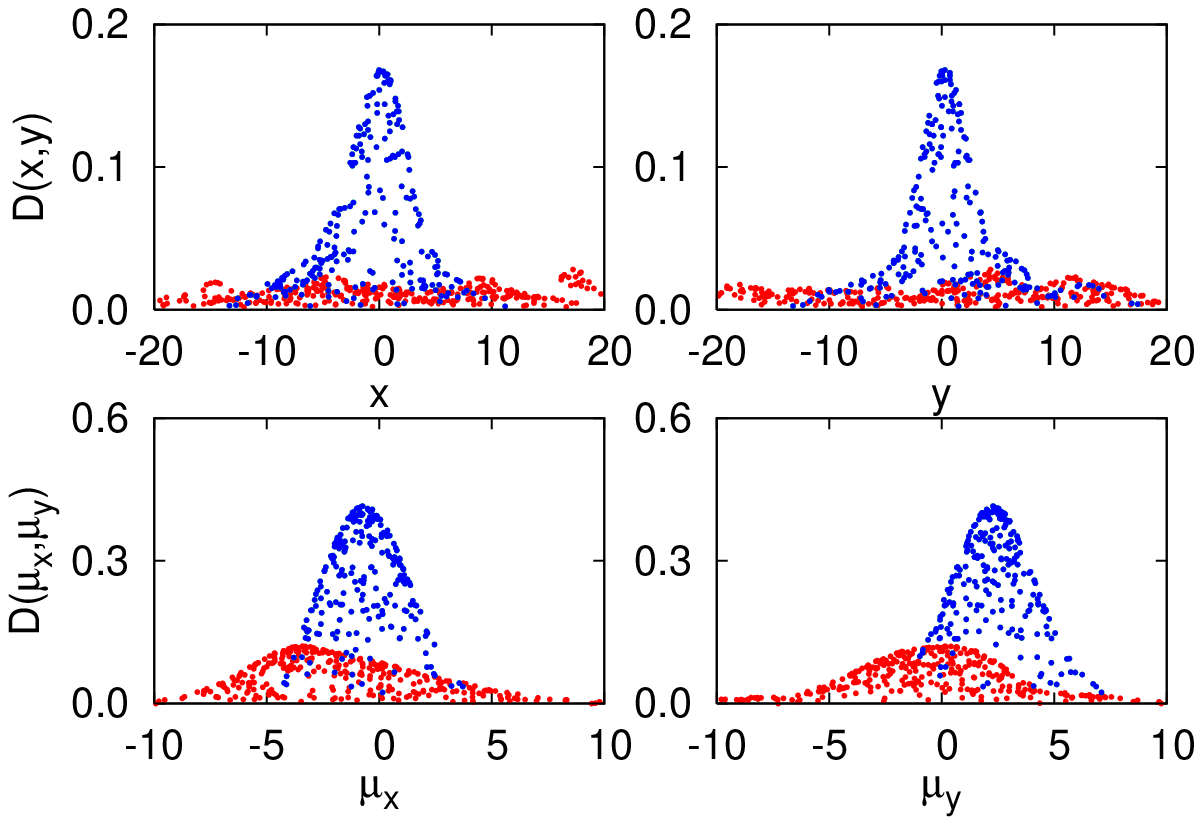}
\caption{Same as Fig.~\ref{pdfic2391}, but for the open
cluster NGC~2194.}
\label{pdfngc2194}
\end{figure}

One advantage of using this method is that the combination
of position and proper motion distributions as membership
criteria, along with the fact that it does not make any
assumption on the underlying distributions, give a higher
degree of flexibility that can make it easier to see the
underlying structure.
Here we show, as illustrative examples, the results for
two different open clusters. In Figures~\ref{ic2391} and
\ref{ngc2194}
we can see both positions and proper motions for the stars
in the region of the open clusters IC~2391 and  NGC~2194,
respectively. We also show, in Figures~\ref{pdfic2391} and
\ref{pdfngc2194},
the corresponding probability density funcions for the same
two clusters. We see that both populations (field stars and
cluster members) have been successfully separated by the
algorithm. The spatial distribution of stars in IC~2391
is more irregular than in NGC~2194, but this is difficult
to see from the spatial distribution
because of the small number of members in
IC~2391. However, the probability density
functions allow a very easy visualization of the spatial
structure. For example, two separate peaks are clearly
visible in IC~2391 located at ($x$,$y$) positions close to
(1,0) and (-0.5,0), and an additional weaker overdensity
close to (0,-2). NGC~2194 exhibits a smoother distribution
in the central region that becomes more
irregular at the border of the cluster.
For example, a small overdensity can be observed close
to position (-5,5). Here we are showing the projected
probability density funcions, obviously the three-dimensional
display allows a better visualization of the cluster structure.

\subsection{Comparison between the parametric
and non-parametric methods}
\label{comparacion}

The methods for discriminating between cluster and field stars
based on the proper motion distributions \citep{Vas58} use
parametric Gaussian functions to represent the corresponding
probability density functions (PDFs). Usually a circular
Gaussian function is assumed for the PDF of the cluster whereas
an elliptical one is adopted for the field. As mentioned in
Section~\ref{salson}, this procedure may present problems if
the underlying PDFs are far from being simple Gaussians, if the
proper motion errors are anisotropic, or if the heterocedastic 
distance between the two stellar populations is small
\citep[see more detailed discussions in][]{Cab85,Cab90,Pla01,Bal04}.
In this case, a suitable option is to apply a non-parametric
discriminating method that determines the PDFs empirically
without a priori assumptions about the profile shapes.
Additionally, even though the underlying PDFs may be well
represented by Gaussians, if the cluster mean proper motion
is very close to the maximum of the field distribution then
the discriminating procedure becomes challengingly difficult.
In fact, the discrimination becomes more difficult as the
statistical distance between both populations decreases.
To increase the statistical distance between cluster and
field it becomes necessary to extend the dimension of the
measurement space, and this is done by including the spatial
coordinates in the non-parametric method used in this study
\citep{Cab90}.

\begin{figure}[th]
\epsscale{.9}
\plotone{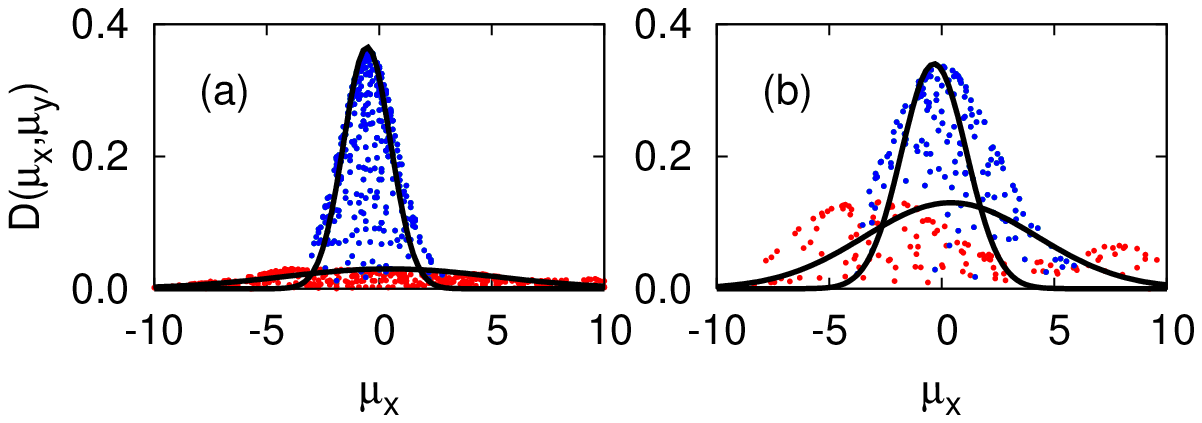}
\caption{Probability density functions in the proper motion
space (in mas/yr) for the stars in the region of (a) M~67 and
(b) NGC~1513. Red and blue circles refer to field and cluster
stars according to the non-parametric method, whereas thick
black solid lines refer to the results of the parametric method.}
\label{ejemplos}
\end{figure}

In order to illustrate (and quantify) these arguments, let
us compare the membership assignments obtained in this work
(Section~\ref{salson}) with those obtained from the classical
parametric method. We have used the algorithm proposed by
\citet{Cab85} which estimates the parameters with a procedure
more simple and efficient than that of \citet{San71}. Moreover, 
the algorithm first identifies outliers in the data in an
objective way, i.e., in a distribution-free way not based
on any previous parameters estimation. This is an important
previous step because outliers make the distribution of field
stars to be flatter than the actual one, modifying the final
probabilities of cluster membership. In order to perform a
better comparison we applied this parametric method to exactly
the same data that we used for the non-parametric method.
As representative examples, Figure~\ref{ejemplos}
shows the resulting PDFs in the proper motion space for two
different open clusters (for a better clarity only the
projection on the coordinate $\mu_x$ is shown). For the
case of M~67 (Figure~\ref{ejemplos}a) the parametric
model finds the position of the cluster centroid at
$\mu_{x,c} = -0.54$ and $\mu_{y,c} = +0.43$ with
$\sigma_c = 1.04$, and the field centroid at
$\mu_{x,f} = +0.46$ and $\mu_{y,f} = +2.22$ with
$\sigma_{x,f} = 4.71$ and $\sigma_{y,f} = 4.53$.
Both parametric and non-parametric PDFs are similar
to each other because cluster and field PDFs are
different enough to allow an adequate separation
of both populations. In fact, 93.67 \% of the stars
in the field of M~67 were assigned to the same
population (cluster member or field star) by both
methods. For NGC~1513 (Figure~\ref{ejemplos}b) the
parametric method finds the cluster centroid at
$\mu_{x,c} = -0.34$ and $\mu_{y,c} = +0.53$ with
$\sigma_c = 1.45$, and the field at
$\mu_{x,f} = +0.41$ and $\mu_{y,f} = +0.24$ with
$\sigma_{x,f} = 3.83$ and $\sigma_{y,f} = 4.06$.
For this case the statistical difference between
both Gaussian PDFs is small in comparison with
M~67 so that, in principle, it is more difficult
to disentangle both populations. The differences
between the parametric and non-parametric PDFs are
more evident and only the 71.4 \% were assigned
to the same class by both methods. The difference
in class assignments arises from the different PDFs
in the proper motion space, but it equally arises
from the fact that the non-parametric method also
uses information from the position space: a star
relatively far from the proper motion centroid might
be classified as a probable cluster member if it was
in a high density region in the corresponding spatial
PDF.

\begin{figure}[th]
\epsscale{.9}
\plotone{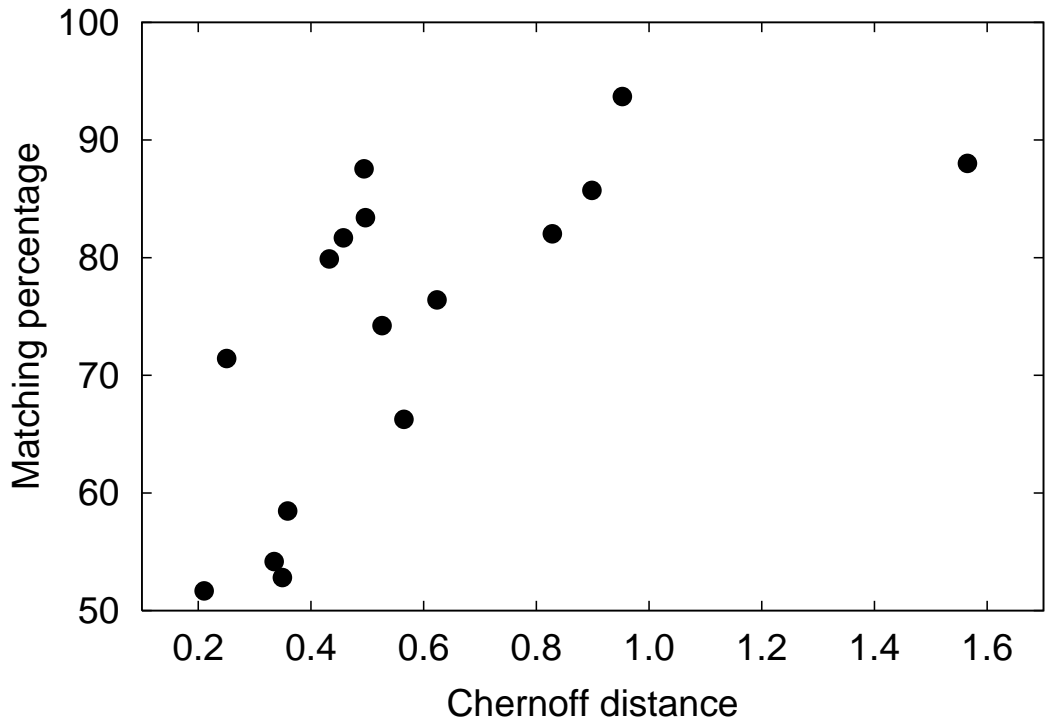}
\caption{Agreement in membership assignation between the
parametric and non-parametric methods as a function of the
Chernoff distance between cluster and field parametric
probability density functions.}
\label{chernoff}
\end{figure}

The statistical separation between any two types of
populations can be described through the Chernoff
probabilistic distance \citep{Che52}, which is a
measure of the difference between two probability
distributions. We have calculated the Chernoff
distance between the two Gaussian PDFs obtained
by the parametric method. This was done for all the
clusters in the sample to quantify the differences 
between the two stellar populations (cluster and
field) in the porper motion space. Figure~\ref{chernoff}
shows the percentage of stars that have obtained the same
assignation (member or non-member) by both methods as a
function of the Chernoff distance. The non-parametric
method used in this work is robust in the sense that
if cluster and field stars can be easily separated in
the proper motion space then the results agree very
well with those of the standar parametric method.
For small Chernoff distances is more difficult to
disentangle both stellar populations only from their
proper motions. In this case, the non-parametric method
has the advantage of using additional information from 
the star positions and then is able to provide a better
discrimination.


\section{Distribution of stars}
\label{resultados}

\begin{figure}[th]
\epsscale{.9}
\plotone{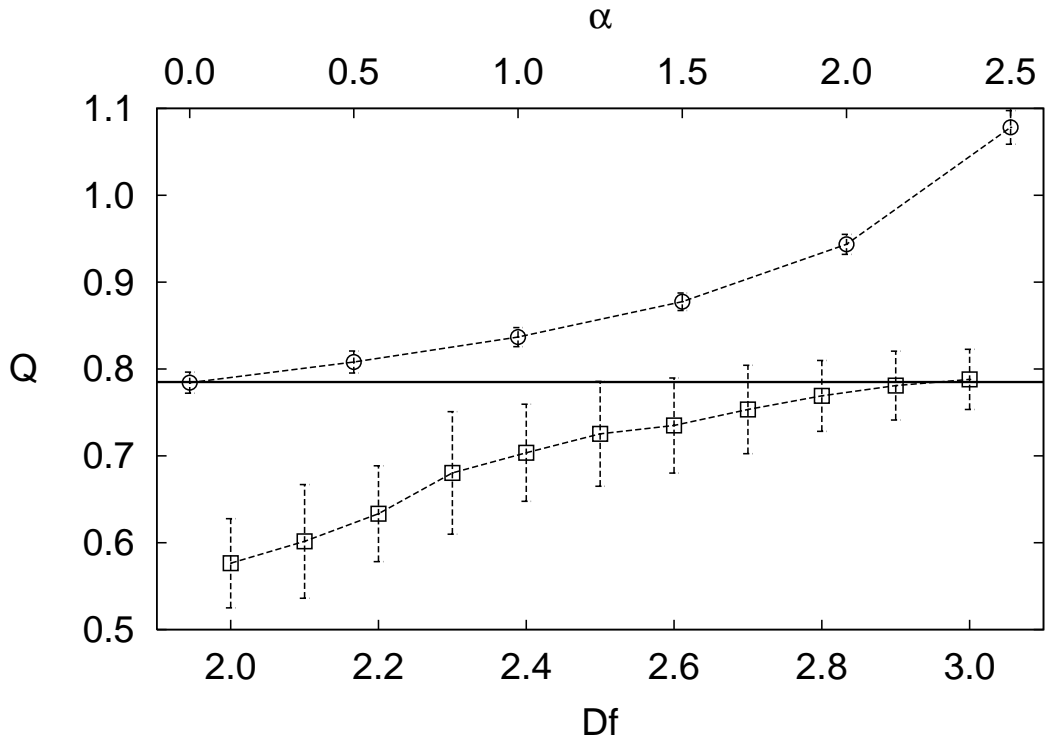}
\caption{Mean values of the parameter $Q$ as a function
of the fractal dimension $D_f$ for projected fractals (open
squares, bottom axis), and as a function of the index $\alpha$
for projected radial profiles (open circles, top axis). The
bars are the corresponding standard deviations. The solid
horizontal line indicates the value $Q=0.785$ for which
both results converge to the homogeneous distribution case
($D_f=3$ and $\alpha=0$).}
\label{teoria}
\end{figure}

We start by using the minimum spanning tree (MST)
technique to analyse the distribution of stars in the
clusters. The MST is the set of straight lines (called
edges) connecting a given set of points without closed
loops, such that the total edge length is minimum.
\citet{Car04} used this technique to study the distribution
of stars in clusters introducing the dimensionless parameter
$Q$. In order to calculate $Q$ we first need to determine the
normalized correlation length $\overline{s}$, i.e. the mean
separation between stars divided by the overall radius of the
cluster. Next, from the MST we determine the normalized mean
edge length $\overline{m}$, i.e. the mean length of the branches
of the tree divided by $(A/N)^{1/2}$ where $A$ is the cluster
area and $N$ the total number of stars. To estimate the area
(and from that the radius) we use the strategy suggested by
\citet{Sch06}, which consists in using the area of the convex
hull, i.e. the minimum-area convex polygon containing the whole
set of data points. Each one of these parameters ($\overline{s}$
and $\overline{m}$) cannot distinguish between a (relatively
smooth) large-scale radial density gradient and a multiscale
(fractal) subclustering. However, \citet{Car04} showed that
the combination $Q=\overline{m}/\overline{s}$ not only is
able to distinguish between radial clustering and fractal
type clustering but can also quantify them. We have generated
two different sets of random three-dimensional distributions
of points: one having a volume density of stars $n$ decreasing
smoothly with the distance from the center $r$ as $n \propto
r^{-\alpha}$ \citep{Car04}, and the other having fractal
patterns according to a recipe that generates distributions
with a well-defined fractal dimension $D_f$ \citep{San08}.
These random simulations were done 50 times, they were
projected on random planes, and then we calculated the 
parameter $Q$ directly from the projected distributions.
The overall results are shown in Figure~\ref{teoria}.
The value $Q=0.785$ (indicated as a horizontal line)
separates radial clustering (open circles) from fractal
clustering (open squares). Moreover, the value of $Q$
itself gives information about the value of $\alpha$
or $D_f$. We have to point out, however, that the
uncertainties for the fractal distributions are rather
large to determine in a precise way $D_f$ from the $Q$
value.

\begin{figure}[th]
\epsscale{.9}
\plotone{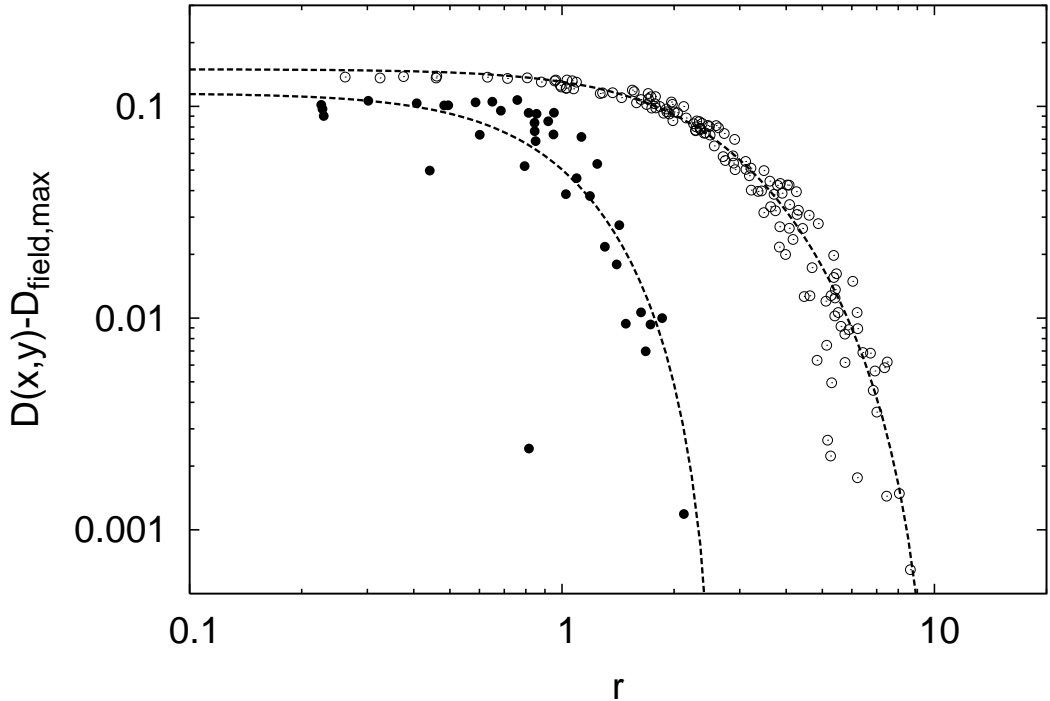}
\caption{Radial density profiles for the members of the
open clusters IC~2391 (solid circles) and NGC~2194 (open
circles). Dashed curves are the King profiles fitted up
to the maximum density of the field stars.}
\label{perfiles}
\end{figure}

We applied this method to the sample of stellar clusters
and the resulting $Q$ values are given in Table~\ref{cumulos}.
Stars in clusters with $Q>0.80$ are distributed following
radial clustering profiles. For a better characterization
of this kind of structure we have fitted \citet{Kin62}
profiles to the radial density distributions of the cluster
members \citep[see][for a discussion on the applicability of
this kind of fit to open clusters]{Hil98}. Before doing the
fit we subtract from the cluster density function the
maximum of the field density function, i.e. we perform the
fit only for the stars in the cluster having probability
densities above the maximum field density.
Figure~\ref{perfiles}
shows the results for the same two example clusters shown
in the previous figures (IC~2391 and NGC~2194). We performed
this fit for all the clusters in our sample, even for the ones
that do not follow smooth profiles. From the best fits we
obtained the core ($R_c$) and tidal ($R_t$) radii. Both
radii are shown in Table~\ref{cumulos}.
Clearly, this fit is unrealistic when the
cluster exhibits a high degree of substructure but, even in
this case, it allows us to estimate the cluster radius
($R_t$) in a homogeneous way for the 16 clusters in the
sample.

Eight of the clusters of our sample (IC~2391, M~34, NGC~581,
NGC~1513, NGC~1647, NGC~1817, NGC~4103, and NGC~6530) have
structure parameter values close to, or below, the threshold
value $Q \simeq 0.80$. These clusters would follow fractal-like
patterns but, as mentioned before, to infer the fractal dimension
from the $Q$ value is quite uncertain. For these clusters, we
choose to estimate the degree of clumpiness by calculating the
correlation dimension ($D_c$). For this we use an algorithm
that estimates $D_c$ in a reliable (precise and accurate) way
\citep{San07a,San08}. The algorithm avoids the usual problems
that arise at relatively large scales (boundary effects) and
small scales (finite-data effects) by using objective and
suitable criteria. Moreover, an uncertainty associated to
each $D_c$ value is estimated using bootstrap
techniques. The application of this algorithm to the eight
clusters having fractal structure yields the results shown
in Table~\ref{Dc_cumulos}.


\section{Discussion}
\label{correlaciones}

\begin{figure}[th]
\epsscale{.9}
\plotone{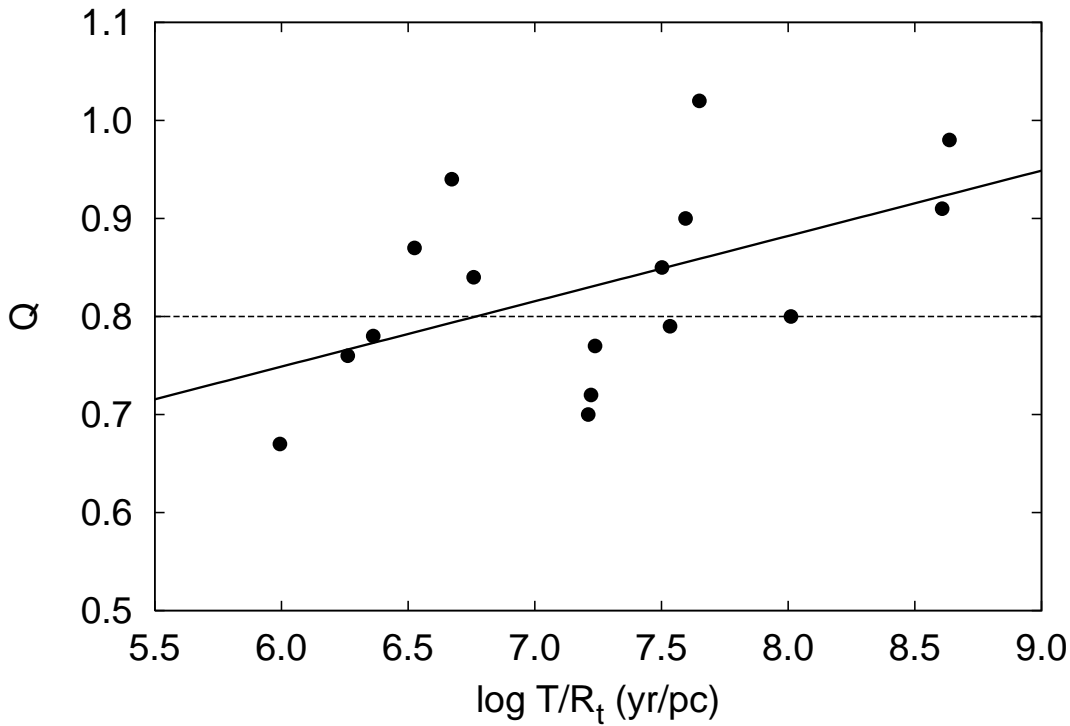}
\caption{Structure parameter $Q$ as a function of the
logarithm of age divided by the tidal radius, which is
nearly proportional to age in crossing time units.
The dashed line at $Q=0.8$ roughly separates radial
from fractal clustering, and the solid line is the
best linear fit.}
\label{correla1b}
\end{figure}

Now we proceed to examine the dependence of $Q$ on the cluster age
in order to compare it with the trend mentioned
by other authors. This kind of dependence has been suggested not
only for stellar clusters \citep{Sch06,Sch08} but also for the
distribution of young stars in the Gould Belt \citep{San07a},
the distribution of young clusters in the solar neighborhood
\citep{Fue06}, and the distribution of stars
\citep{Bas09,Gie08,Ode08} and HII regions \citep{San08} in
external galaxies. A slight positive trend
is apparent when we plot $Q$ versus $\log(T)$, i.e. fractal
clusters tend to be younger than clusters having radial density
profiles. However, the statistical analysis indicates that
there is no significant correlation between this structure
parameter and cluster age, neither for the full sample nor
for the fractal clusters and density profile clusters
considered individually.
From simple arguments one would expect that $Q$ increases
with time {\it for each cluster}. Gravitationally unbound
cluster will tend to nearly homogeneous distributions
($Q=0.79$, $D_c=2.0$) because of the dispersal of stars,
whereas self-gravity will lead to more centrally peaked 
distributions in bound clusters. It could take several
crossing times to reach an equilibrium state and/or to
eliminate the original distribution \citep{Bon98,Goo04},
although maybe it could take only a crossing time \citep{Bas09}.
The typical crossing time in open clusters is of the order of
$10^6$ years \citep{Lad03} but, assuming nearly the same typical
velocity dispersion, the crossing time is roughly proportional
to the cluster size. Let us consider the new variable, $T/R_t$
(in yr/pc), which is proportional to time measured in crossing
time units. In this case we do observe the correlation
\begin{displaymath}
Q = (0.07 \pm 0.03) \log (T/R_t) + (0.35 \pm 0.21)\ \ \ ,
\end{displaymath}
which is significant at 96\% confidence level.
This result is shown in Figure~\ref{correla1b}.

Previous detailed studies on the fractal properties of projected
distributions of points \citep{San07a,San08} have shown that the
uncertainty associated with $D_c$ depends on the number of available
data. Moreover, when the number of data points is too low
($N \lesssim 200$) a bias in the mean $D_c$ values is produced.
We performed a similar analysis for the parameter $Q$ using the
simulated fractals. We verified that the mean measured value of
$Q$ tends to be overestimated if $N \lesssim 200$, and the bias
was higher as the fractal dimension (and therefore $Q$) decreased.
For the extreme case studied here ($D_f = 2$), the maximum difference
between the mean value of $Q$ for well-sampled point sets (namely
$Q=0.576$, see Figure~\ref{teoria}) and fractals having $N \sim
200$ data points was $\Delta Q \simeq 0.06$. The important point
here is that if this kind of bias is present in our results, then
the correlation shown in Figure~\ref{correla1b} might be reinforced.

\begin{figure}[th]
\epsscale{.9}
\plotone{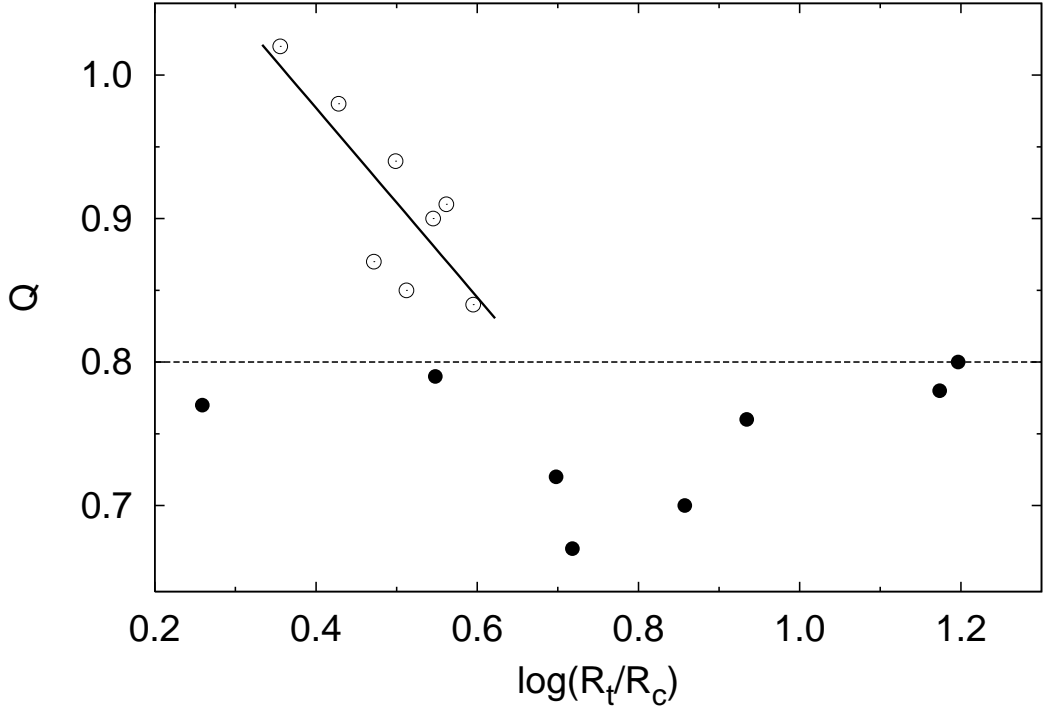}
\caption{Structure parameter as a function of
the concentracion parameter of the King model. The dashed line
($Q=0.8$) roughly separates clusters with well-defined radial
density profiles (open circles) and clusters with substructures
(filled circles). The solid line is the best linear fit for the
upper subsample.}
\label{QvsLogC}
\end{figure}

The structure parameter is shown in Figure~\ref{QvsLogC}
as a function of the concentration parameter of the King model.
Interestingly, the behaviors of the subsamples $Q > 0.8$ and
$Q \leq 0.8$ are clearly differentiated. $Q$ correlates strongly
with the concentration for cluster with well-defined radial 
density profiles, the best linear fit (solid line in 
Fig.~\ref{QvsLogC}) being
\begin{displaymath}
Q = -(0.66 \pm 0.20) \log (R_t/R_c) + (1.24 \pm 0.10)\ \ \ ,
\end{displaymath}
with a confidence level greater than 98\%. Otherwise, the
fractal-like subsample does not show any correlation at all.

\begin{figure}[th]
\epsscale{.9}
\plotone{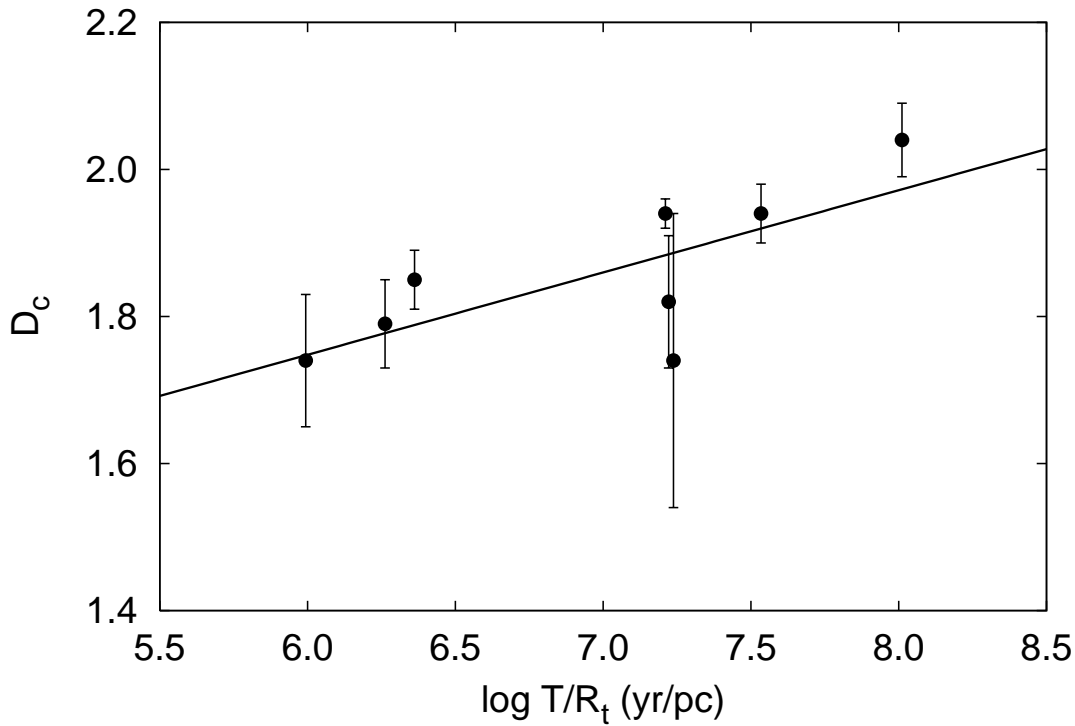}
\caption{Calculated correlation dimension as a function of
the logarithm of age divided by the tidal radius. The solid
line is the  best linear fit.}
\label{correla2b}
\end{figure}

As we have seen, there seems to be some evidence that young
clusters tend to distribute their stars following fractal
patterns whereas older clusters tend to exhibit centrally
concentrated structures. But this is only an overall trend.
Note, for example, that NGC~1513 and NGC~1647 have both
$Q \sim 0.7$ with ages of $T \gtrsim 100$ Myr. The advantage
of analyzing the clustering properties via the correlation
dimension is that, apart from {\it directly} measuring
the fractal dimension, the assignment of an associated uncertainty
allows us to know the reliability of each measurement. The results
of $D_c$ for the clusters having $Q\lesssim 0.8$ are shown in
Table~\ref{Dc_cumulos}. The best linear fit between the fractal 
dimension and the age is:
\begin{displaymath}
D_c = (0.14 \pm 0.05) \log (T) + (0.77 \pm 0.39)\ \ \ ,
\end{displaymath}
significant at a confidence level of 97\%.
If we use $T/R_t$ instead $T$ the fit becomes:
\begin{displaymath}
D_c = (0.11 \pm 0.04) \log (T/R_t) + (1.08 \pm 0.30)\ \ \ ,
\end{displaymath}
significant at a level of 96\%. This last fit is shown in
Figure~\ref{correla2b},
where we can see that the correlation looks very good by eye.
The point farthest from the best-fit line is the cluster
IC~2391, which has the smallest number of members ($N_s=62$)
and the largest uncertainty in $D_c$ ($0.2$). If the result
for this cluster is biased, then the fractal dimension should
be higher than the value reported here and the correlation
should be even stronger. An important aspect to be mentioned
is that there exist stellar clusters as old as $\sim 100$ Myr
that have not totally destroyed their clumpy substructure.
This is a particularly meaningful result that gives some
observational support to recent simulations of the dynamical
evolution of young clusters \citep{Goo04}.

We have already mentioned that converting from two-dimensional
to three-dimensional fractal dimensions increases
the associated uncertainties. However, it is
interesting to note that according to our previous works
\citep{San07a,San08}, clusters with the smallest correlation
dimensions ($D_c=1.74$) would have three-dimensional fractal
dimensions around $D_f \sim 2.0$. This values is considerably
smaller than the average value $D_f \simeq 2.6-2.7$ estimated
for the interstellar medium in recent studies \citep{San05,San07b}.
Perhaps the development of some kind of substructure in initially
more homogeneous clusters observed in some simulations could explain
this difference, although some coherence in the initial velocity
dispersion would be necessary \citep{Goo04}. Another plausible
explanation is that this difference is a consequence of a more
clustered distribution of the densest gas at the smallest spatial
scales in the molecular cloud complexes, according to a multifractal
scenario for the interstellar medium \citep{Cha01,Tas07}.
The problem is complex because it depends on: (a) the initial 
distribution of gas and dust in the parent cloud, (b) the way and
degree in which this information is transferred to the new-born stars,
and (c) how, and how fast, this initial star distribution evolves.
Each one of these factors will depend to a greater or lesser extent 
on the involved physics and environmental variables. These points 
clearly require more investigation.


\section{Conclusions}
\label{conclusiones}

We have characterized quantitatively the distribution of stars
in a relatively large sample of open clusters (a total of 16) 
spanning a wide range of ages. Membership probabilities were
obtained by applying a non-parametric method that does not make
any assumption on the underlying star distribution. This is a
crucial point to avoid possible bias introduced by the cluster
member selection process.
We found evidence that stars in young clusters
tend to be distributed following clustered, fractal-like patterns,
whereas older clusters tend to exhibit radial star density profiles.
This result supports the idea that stars in new-born cluster likely
follow the fractal patterns of their parent molecular clouds, and that
eventually evolve toward more centrally concentrated structures
\citep[see also][]{Sch06}. However, we have also obtained
some other interesting results: (a) there exists a
strong correlation between the structure parameter $Q$ and the
concentration parameter of the King model $\log (R_t/R_c)$ for
the clusters with well-defined radial density profiles, (b)
clusters as old as $\sim 100$ Myr can exhibit a high degree of
spatial substructure, and (c) there is a significant correlation
between fractal dimension and age for the cluster with fractal
distribution of stars. Additionally, we find that the smallest
values of the corresponding three-dimensional fractal dimensions
are $D_f \sim 2.0$, which is considerably smaller than the value
$D_f \simeq 2.6-2.7$ estimated for the average interstellar gas
distribution. If this is a general result, then some further
explanation would be required.

\acknowledgments
We want to thank the referee for his/her
comments which improved this paper.
This research has made use of the VizieR database (operated
at CDS, Strasbourg, France), the WEBDA database (operated
at the Institute for Astronomy of the University of Vienna),
and the  NASA's Astrophysics Data System.
We acknowledge financial support from MICINN of Spain through grant
AYA2007-64052 and from Consejer\'{\i}a de Educaci\'on y Ciencia
(Junta de Andaluc\'{\i}a) through TIC-101. N.S. is supported by
a post-doctoral JAE-Doc (CSIC) contract.



\input{tab1.tex}
\input{tab2.tex}

\end{document}

%% file: tab1.tex
\begin{deluxetable}{lcrrrcccc}
\tablecolumns{9}
\tablewidth{0pt}
\tablecaption{Properties of the clusters in the sample\label{cumulos}}
\tablehead{
\colhead{Name} &
\colhead{$\log(T)$} &
\colhead{$D$} &
\colhead{$N_d$} & 
\colhead{$N_s$} &
\colhead{$Q$} &
\colhead{$R_c$} &
\colhead{$R_t$} &
\colhead{Ref.}
}
\startdata
IC 2391  & 7.661 &  175 & 6847 &   62 & 0.77 & 1.46 &  2.65 & (1) \\
M 67     & 9.409 &  908 & 1046 &  354 & 0.98 & 2.21 &  5.92 & (2) \\
M 11     & 8.302 & 1877 &  872 &  289 & 1.02 & 1.98 &  4.49 & (3) \\
M 34     & 8.249 &  499 &  630 &  181 & 0.80 & 0.11 &  1.73 & (4) \\
NGC  188 & 9.632 & 2047 & 7771 & 1459 & 0.91 & 2.90 & 10.57 & (5) \\
NGC  581 & 7.336 & 2194 & 2387 &  526 & 0.76 & 1.38 & 11.86 & (6) \\
NGC 1513 & 8.110 & 1320 &  332 &  156 & 0.72 & 1.55 &  7.73 & (7) \\
NGC 1647 & 8.158 &  540 & 2220 &  683 & 0.70 & 1.23 &  8.86 & (8) \\
NGC 1817 & 8.612 & 1972 &  810 &  277 & 0.79 & 3.39 & 11.97 & (9) \\
NGC 1960 & 7.468 & 1318 & 1190 &  311 & 0.87 & 2.96 &  8.77 & (10) \\
NGC 2194 & 8.515 & 3781 & 2233 &  228 & 0.85 & 3.17 & 10.31 & (10) \\
NGC 2548 & 8.557 &  769 &  501 &  168 & 0.90 & 2.61 &  9.16 & (11) \\
NGC 4103 & 7.393 & 1632 & 4379 &  799 & 0.78 & 0.72 & 10.74 & (12) \\
NGC 4755 & 7.216 & 1976 &  384 &  196 & 0.94 & 1.11 &  3.50 & (12) \\
NGC 5281 & 7.146 & 1108 &  314 &   80 & 0.84 & 0.62 &  2.44 & (12) \\
NGC 6530 & 6.867 & 1330 &  364 &  145 & 0.67 & 1.43 &  7.47 & (13)
\enddata
\tablerefs{(1) \citet{Pla07}; (2) \citet{Zha93}; (3) \citet{Su98};
(4) \citet{Jon96}; (5) \citet{Pla03}; (6) \citet{San99}; (7) \citet{Fro02};
(8) \citet{Gef96}; (9) \citet{Bal04}; (10) \citet{San00}; (11) \citet{Wu02};
(12) \citet{San01}; (13) \citet{Zha06}.}
\end{deluxetable}

%% file: tab2.tex
\begin{deluxetable}{lc}
\tablecolumns{2}
\tablewidth{0pt}
\tablecaption{Calculated Correlation Dimensions\label{Dc_cumulos}}
\tablehead{ \colhead{Name} & \colhead{$D_c$} }
\startdata
IC~2391  & $1.74 \pm 0.20$ \\
M~34     & $2.04 \pm 0.05$ \\
NGC~581  & $1.79 \pm 0.06$ \\
NGC~1513 & $1.82 \pm 0.09$ \\
NGC~1647 & $1.94 \pm 0.02$ \\
NGC~1817 & $1.94 \pm 0.04$ \\
NGC~4103 & $1.85 \pm 0.04$ \\
NGC~6530 & $1.74 \pm 0.09$
\enddata
\end{deluxetable}